\documentclass[aps,prl,reprint,superscriptaddress]{revtex4-1}
\usepackage{graphicx}
\usepackage{dcolumn}
\usepackage{bm}
\usepackage{color}
\usepackage{sidecap}

\begin{document}

\title{Nonreciprocal charge transport in the titanium sesquioxide heterointerface superconductor}

\author{Peng Dong} 
\affiliation{ShanghaiTech Laboratory for Topological Physics \& School of Physical Science and Technology, ShanghaiTech University, Shanghai 201210, China}

\author{Lijie Wang} 
\affiliation{State Key Laboratory of Surface Physics and Department of Physics, Fudan University, Shanghai 200433, China}

\author{Guanqun Zhang} 
\affiliation{State Key Laboratory of Surface Physics and Department of Physics, Fudan University, Shanghai 200433, China}

\author{Jiadian He}
\author{Yiwen Zhang}
\author{Yifan Ding}
\author{Xiaohui Zeng}
\author{Jinghui Wang}
\author{Xiang Zhou}
\affiliation{ShanghaiTech Laboratory for Topological Physics \& School of Physical Science and Technology, ShanghaiTech University, Shanghai 201210, China}

\author{Yueshen Wu}
\email{wuysh@shanghaitech.edu.cn}
\affiliation{ShanghaiTech Laboratory for Topological Physics \& School of Physical Science and Technology, ShanghaiTech University, Shanghai 201210, China}

\author{Wei Li}
\email{w\_li@fudan.edu.cn}
\affiliation{State Key Laboratory of Surface Physics and Department of Physics, Fudan University, Shanghai 200433, China}

\author{Jun Li}
\email{lijun3@shanghaitech.edu.cn}
\affiliation{ShanghaiTech Laboratory for Topological Physics \& School of Physical Science and Technology, ShanghaiTech University, Shanghai 201210, China}

\date{\today}

\begin{abstract}
Nonreciprocal charge transport in heterostructural superconductors exhibits appealing quantum physical phenomena and holds the promising potential for superconducting circuits applications. Realizing a nonreciprocity is, however, fundamentally and technologically challenging, as it requires a material structure without a centre of inversion, which is scarce among superconducting materials. Here, we report an evidence of helical superconductivity, in which the Rashba spin-orbit coupling induces momentum-dependent superconducting gap in the inversion symmetry breaking heterointerface superconductor consisting of Mott insulating Ti$_2$O$_3$ and polar semiconducting GaN. Remarkably, the nonlinear responses emerge in the superconducting transition regime, when the magnetic field is precisely aligned in-plane orientations perpendicular to the applied current. In particular, the observed nonreciprocal supercurrent is extremely sensitive to the direction of the magnetic field for 0.5 degree, suggestive of a crossover from a symmetry breaking state to a symmetric one. Our finding not only unveils the underlying rich physical properties in heterointerface superconductors, but also provides an exciting opportunity for the development of novel mesoscopic superconducting devices.

\end{abstract}

\maketitle

Symmetry breaking plays an essential role in the emergence of superconductivity and influences many of its physical properties in a profound way, which provides a fundamental understanding of the Cooper pairs formation in superconductivity~\cite{Valentini,Hartke,Wang2,Bovzovic,Kirtley,Ji,Hamidian}. In conventional superconductors~\cite{Bardeen}, a condensate of Cooper pairs spontaneously breaks only the~\emph{U}(1) gauge symmetry. In unconventional superconductors~\cite{Kirtley,Sigrist,Wux,Xu2,Roppongi,Yuan,xue2022}, however, additional symmetries will be further broken, giving rise to enormous novel physical behaviors and the possibility of multiple superconducting phases. For example, broken rotational symmetry in high-$\emph{T}$$_c$ cuprate superconductors could exhibit an anisotropic~\emph{d}-wave pairing symmetry~\cite{Shen,Tsuei}, broken time-reversal symmetry in superconductivity could induce spin-triplet Cooper pairs~\cite{Hillier,Shang,Ning}, and broken lattice inversion symmetry in non-centrosymmetric compounds could result in the coexistence of spin-singlet and spin-triplet pairings~\cite{Hillier,Shang}. Strikingly, two-dimensional heterointerface superconductors possess a natural inversion symmetry breaking, the existence of the interfacial electric field leads to strong spin-orbit coupling (SOC), resulting in the mixed-parity superconductivity with an admixture of~\emph{s}-wave and~\emph{p}-wave pairings~\cite{Kozii,Zhang,zhang2023}, a candidate platform for realizing Majorana modes~\cite{Potter}. Consequently, it is pivotal to reveal the emergent fascinating and non-trivial superconducting properties at the heterointerfaces with inversion symmetry breaking and develop the next generation quantum technologies.

To illustrate appealing physical properties reflecting an inversion symmetry breaking in lattice and electronic structures of the two-dimensional superconductors, nonreciprocal supercurrent (NSC), which emerges only if both inversion and time-reversal symmetries are broken~\cite{Ando,Narita,Bauriedl,Wu}, has recently played a pivotal role in electronic transport~\cite{Bauriedl,Wakatsuki,Itahashi,Itahashi1,Lyu,WuYueshen}.
A typical electronic band structure with Rashba-like SOC could be responsible for the NSC in an interfacial superconductivity of Bi$_2$Te$_3$/FeTe~\cite{Kenji} and gate-induced two-dimensional superconductivity on the surface of SrTiO$_3$~\cite{Itahashi}. The nonlinearity of electric resistance $R_{\text{xx}}^{2\omega}$ that depends on the applied current $\textbf{\emph{I}}$ and the external magnetic field $\textbf{\emph{B}}$ is thus phenomenologically expressed as: $R_{\text{xx}}^{2\omega}$=\emph{R}$_0$$\gamma$(\textbf{\emph{B}}$\times$\textbf{\emph{P}})$\cdot$\textbf{\emph{I}}, where \textbf{\emph{P}} is a unit vector which characterizes the axis of the nonreciprocal effect \cite{Kenji}. Moreover, it is interesting to elucidate the underlying rich physical properties of Rashba-like systems that have been proposed theoretically, including helical and chiral superconductivity. The former one breaks the inversion symmetry in the presence of magnetic field, while the latter one further spontaneously breaks the time reversal symmetry. The origin of these quantum states, however, remain under intense scrutiny in experiments. Alternatively, using state-of-the-art heterostructure engineering, we have developed an interfacial superconductivity of Ti$_2$O$_3$/GaN as a following of a striking quantum metallic state~\cite{Wang,zhang2022}. Due to the presence of the strong interfacial coupling between the Mott insulator Ti$_2$O$_3$ and the polar semiconductor GaN, it would be crucial to exploit the NSC behaviors inherent to the superconducting pair symmetry, offering innovative understanding of the underlying nature of the exotic heterointerface superconductivity.

\begin{figure*}[!htbp]
	\includegraphics[width=0.9\textwidth,clip]{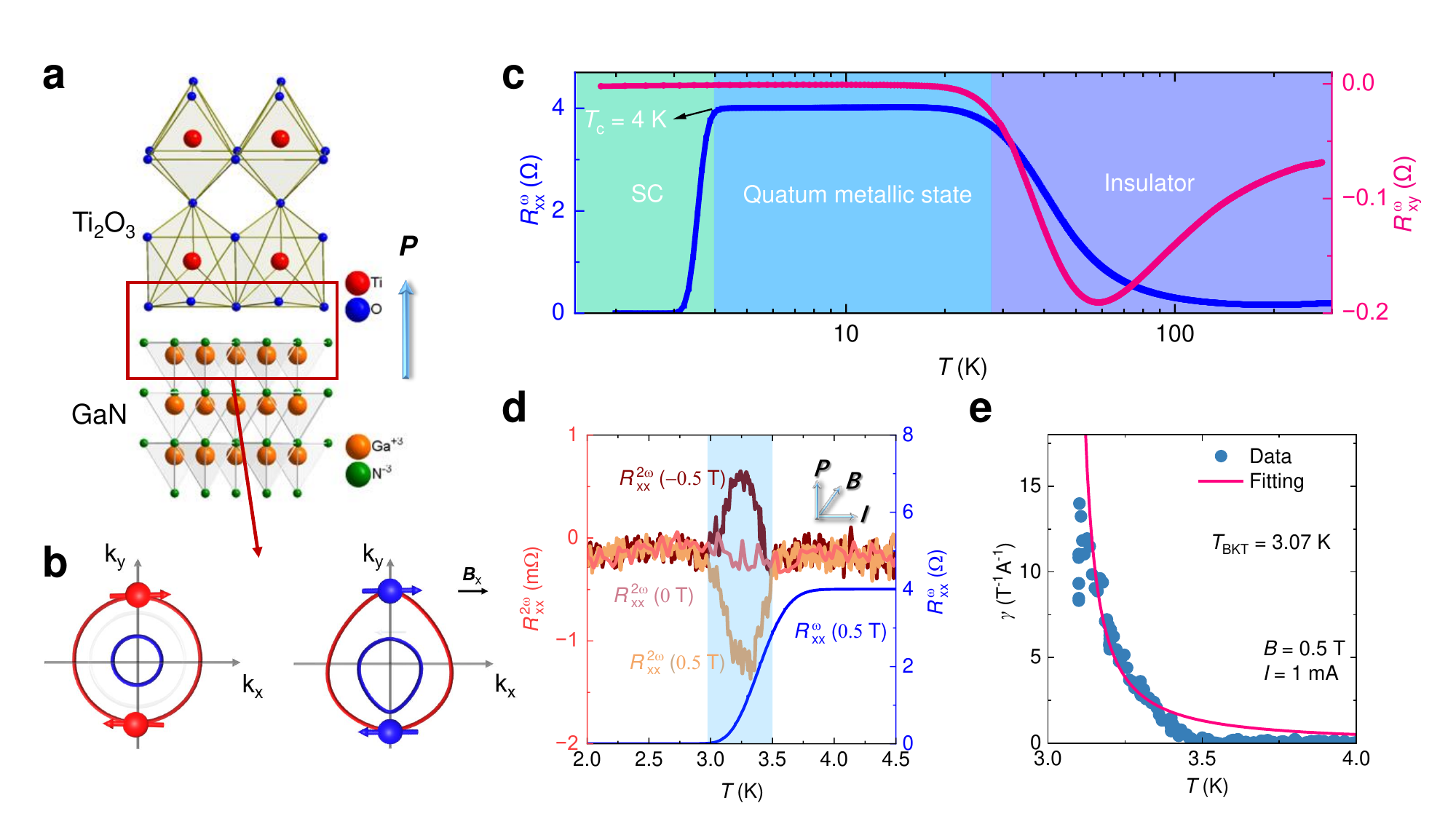}
	\caption{(color online). \textbf{Transport measurement and nonreciprocity in Ti$_2$O$_3$/GaN heterostructure.} (a) Side view of titanium sesquiodide heterointerface and the direction of polarity ($P$) in real space in Ti$_2$O$_3$/GaN. (b) Nonreciprocal charge transport in Rashba-system due to interface polarization. (c) Temperature dependent longitudinal resistance $R_{\text{xx}}^{\omega}$ (blue) and transverse resistance $R_{\text{xy}}^{\omega}$ (pink, under magnetic field 5 T) resistance in Ti$_2$O$_3$/GaN heterostructure. (d) Longitudinal resistance (blue) $R_{\text{xx}}^{\omega}$ and second harmonic resistance $R_{\text{xx}}^{2\omega}$ (orange) as a function of temperature in the low temperature region, where the magnetic field and current are applied along the $y$-axis and $x$-axis, respectively. $R_{\text{xx}}^{2\omega}$ shows just a background noise under zero field, while exhibits obvious signal at the superconducting transition region under field of 0.5 T, and particularly, the sign of $R_{\text{xx}}^{2\omega}$ depends on field as the relationship $R_{\text{xx}}^{2\omega}$=$R_0$$\gamma$(\textbf{\emph{B}}$\times$\textbf{\emph{P}})$\cdot$\textbf{\emph{I}}. (e) Nonreciprocal strength $\gamma$ as a function of temperature in superconductor fluctuation region.
	}
\end{figure*}

In this Letter, we delve into the nonreciprocity of the superconductivity at Ti$_2$O$_3$/GaN heterointerface. The temperature and magnetic field dependence of second harmonic resistance $R_{\text{xx}}^{2\omega}$ shows the value of nonreciprocal strength $\gamma$ as large as 12 $\text{A}^{-1}\text{T}^{-1}$. 
Particularly, only the in-plane magnetic field can induce NSC, and a slight tile of the magnetic field above $\pm 0.5^{\circ}$ will completely suppress the nonreciprocity. Such sensitively angular dependent phenomenon puts a strong confinement to the origin of the symmetry breaking, which suggests a magnetic field induced switching of pairing symmetry.

The Ti$_2$O$_3$ film was grown on a (0001)-oriented GaN substrate by using pulsed laser deposition in an ultrahigh vacuum chamber with a base pressure of 10$^{-9}$ Torr as described in our previous report~\cite{Wang}. The superconducting interface developed between Mott insulator Ti$_2$O$_3$ and wide-band gap GaN attributing to charge transfer from the polar GaN substrate to the Ti$_2$O$_3$ layer and the existence of oxygen vacancies, as shown in Fig 1(a). The polarity perpendicular to the interface also leads to Rashba-type SOC, and then resulting in a band splitting along the \textbf{\emph{k}}$_x$ axis in \textbf{\emph{k}} space. When a magnetic field is applied along the \textbf{\emph{k}}$_x$ direction, Fermi pocket will be shifted along the \textbf{\emph{k}}$_y$ direction due to the Zeeman energy as illustrated in Fig 1(b). As a consequence, the interplay between two helical bands with opposite modulation vectors will give rise to a nonreciprocal charge transport when the current flows along the \textbf{\emph{k}}$_x$ direction due to formation of helical state \cite{Ilic,Daido,Yuan}.

\begin{figure*}[!htbp]
	\includegraphics[width=0.9\textwidth,clip]{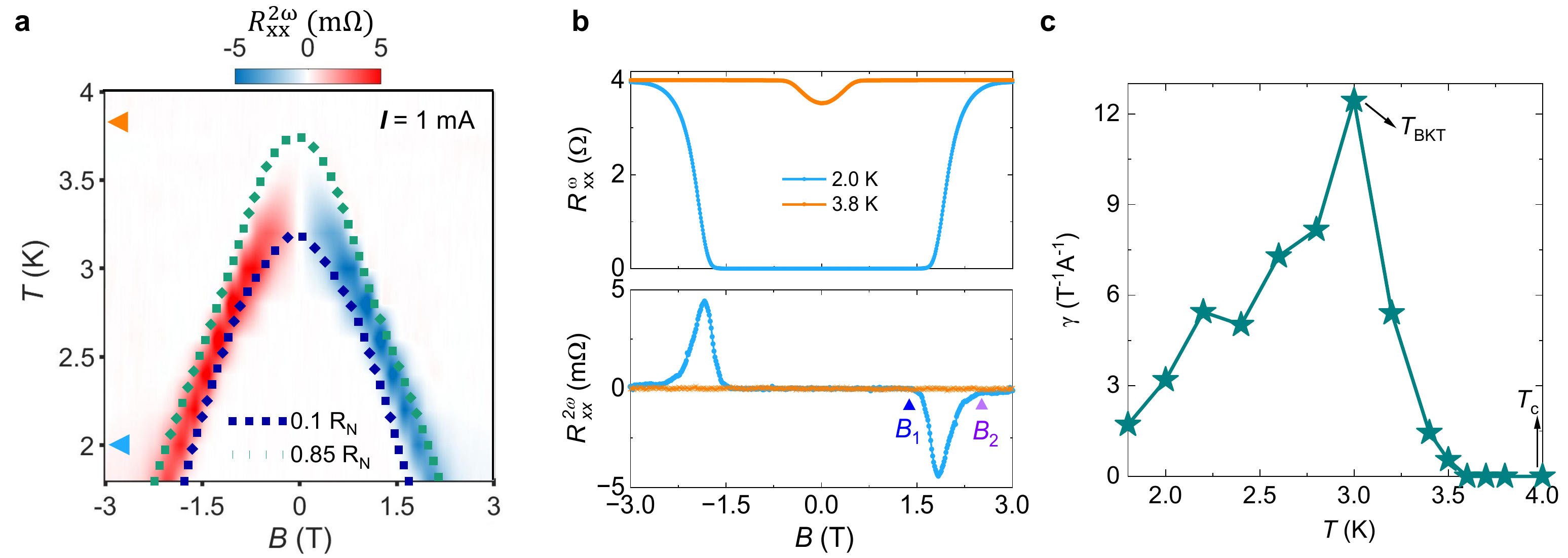}
	\caption{(color online). \textbf{Nonreciprocal charge transport as a function of magnetic field in different temperature in Ti$_2$O$_3$/GaN heterostructure.} (a) The mapping of $R_{\text{xx}}^{2\omega}$ as a function of magnetic field in temperature range from 1.8 K to 4 K, which is antisymmetric in regard to positive or negative magnetic field. Dash line indicate 0.1$R_{\text{N}}$ and 0.85$R_{\text{N}}$ in $R_{\text{xx}}^{\omega}$, where $R_{\text{N}}$ is the normal state resistance at 4.7 K. (b) First and second harmonic resistance at 2 K and 3.8 K, corresponding to the triangle position marked in (a). (c) Temperature dependent nonreciprocal strength $\gamma$=2$\cdot$$R_{\text{xx}}^{2\omega}$/{\emph{B}\emph{I}$R_{\text{xx}}^{\omega}$} at the normal state and superconducting fluctuation region.
	}
\end{figure*}

To probe the nonreciprocal transport, a hall-bar device configuration was defined with the channel length of 300 ${\mu}$m and with the width of 50 ${\mu}$m by standard photolithography technique using photoresist (AZ5214) and the Ar ion beam milling (commercial Intlvac system) process to selectively remove excess materials. Nonreciprocity of the device was detected by a standard lock-in technique with four-probe geometry. All electrical measurements were performed in a Quantum Design Physical Property Measurement System (PPMS) with commercial sample rotators. The first and second harmonic resistance were defined as $R_{\text{xx}}^{\omega}$ = $\sqrt{2}${$V_{\text{xx}}^{\omega}$}/{\emph{I}} and $R_{\text{xx}}^{2\omega}$ = $\sqrt{2}${$V_{\text{xx}}^{2\omega}$}/{\emph{I}},where $I$ is the amplitude of the sinusoidal AC current source, and $V_{\text{xx}}^{\omega}$ and $V_{\text{xx}}^{2\omega}$ are the amplitude of first and second harmonic voltage measured by lock-in amplifier. A phase shift of $\pi$/2 was added to the second harmonic signal. In the Rashba system, the second harmonic resistance should follow the relation of
$R_{\text{xx}}^{2\omega}$=\emph{R}$_0$$\gamma$(\textbf{\emph{B}}$\times$\textbf{\emph{P}})$\cdot$\textbf{\emph{I}}, according to the relation between current, magnetic field and polarity. We then are able to distinguish the role of the Rashba-type SOC by examining this relation.

In Fig 1(c), as temperature decreases, longitudinal resistance $R_{\text{xx}}^{\omega}$ increases continuously until 27 K, and then it reveals a wide range of resistance platform until a superconducting state below critical temperature \emph{T}$_c$ of 4 K. Transverse resistance $R_{\text{xy}}^{\omega}$ (pink) under field of 1 T reveals a compensation of the carriers below 40 K, in agreement with the previous work \cite{Wang}. The second harmonic longitudinal resistance $R_{\text{xx}}^{2\omega}$ is observed at the superconducting transition (blue region) by applying an in-plane magnetic field of 0.5 T as schematically illustrated in the inset in Fig 1(d). The sign of $R_{\text{xx}}^{2\omega}$ can be reversed when the magnetic field is tuned from 0.5 T to -0.5 T, which is congruous with the field dependent nonreciprocal signal. We does not found nonreciprocity at 0 T, indicating no chirality in the system.

Since the superconducting transition in 2D limit is governed by Berezinskii-Kosterlitz-Thouless theory (BKT) theory and thermal fluctuation above and below the mean field critical temperature \emph{T}$_{c0}$, we then carefully extracted BKT transition temperature \emph{T}$_{\text{BKT}}$ and \emph{T}$_{c0}$ by fitting the $R-T$ curve with Halperin-Nelson formula (red dashed line)
	$ R_{xx} = a\cdot e^{b\sqrt{\left ( T_{c0}-T \right ) / \left ( T-T_{\text{BKT}}   \right )  } } $
where $a$= 4 $\Omega$ is normal state resistance, $b$= -2.4 is dimensionless constant, and $T_{\text{BKT}}$ = $3.07$ $\pm$ $0.01$ K, $T_{\text{c0}}$= 3.4 K  at  0.5 T (Supplemental Material Fig S1). The nonreciprocal charge transport appears in the range of 3 K to 3.5 K, indicating the origin of BKT transition or paraconductivity. We calculate the coefficient $\gamma$ representing the strength of the magnetochiral anisotropy, according to the definition of $\gamma$=2$\cdot$$R_{\text{xx}}^{2\omega}$/{\emph{B}\emph{I}$R_{\text{xx}}^{\omega}$}.
Fig 1(e) shows that $\gamma$ increases as the temperature decreases. The data are fitted by formula $\gamma =\text{A} \left ( T-T_{\text{BKT}}  \right ) ^{-\text{B}}$, where $\text{A}$=$0.46$ $\pm$ $0.05$ $\text{A}^{-1}\text{T}^{-1}\text{K}^{3/2}$ and $\text{B}$=$1.23$ $\pm$ $0.05$, which agrees with the BKT transition.

We then check the magnetic field and temperature dependence of $R_{\text{xx}}^{2\omega}$. Figure 2(a) shows the mapping of the temperature depdendent $R_{\text{xx}}^{2\omega}$, where most of the nonreciprocal signals exist in the contour of $R$ = 0.1 $R_{\text{N}}$ and $R$ = 0.85 $R_{\text{N}}$ (dashed lines), where $R_{\text{N}}$ is the normal resistance at 4.5 K, and the raw data are given in Fig. S3. The temperature dependence of maximum $R_{\text{xx}}^{2\omega}$ in $R-H$ curves and $\gamma$ at maximum $R_{\text{xx}}^{2\omega}$ are calculated in Fig 2(c). The maximum value of the $\gamma$ peak at 3 K is about 12 $\text{A}^{-1}\text{T}^{-1}$, which is larger than those reported in other non-superconducting systems such as rashba semiconductor GeTe ($\sim$ $10^{-3}$ $\text{A}^{-1}\text{T}^{-1}$) \cite{Li}, chiral organic materials ($\sim$ $10^{-2}$ $\text{A}^{-1}\text{T}^{-1})$ \cite{Pop}, BiTeBr ($\sim$ $1$ $\text{A}^{-1}\text{T}^{-1})$ \cite{Ideue}, Bi$_2$Te$_3$/FeTe interfacial superconductor \cite{Kenji}, and the noncentrosymmetric oxide interfaces LaAlO$_3$/SrTiO$_3$ \cite{Choe}, while less than the noncentrosymmetric layered superconductors \cite{Wakatsuki,Zhang}. It worth noting that the supression of the effect at lower temperature is consistent with the theroy calculation because one helical band begins to dominate the other as the critical field increases \cite{Ilic}.

Figure 3(a) shows the magnetic field dependence of $R_{\text{xx}}^{2\omega}$ for various value of \emph{I}, which increases as current up to 4 mA, follow by a drop as futher increasing current. We extract the maximum value of $R_{\text{xx}}^{2\omega}$ as a function of current and plot in Fig. 3(b). $R_{\text{xx}}^{2\omega}$ exhibits a linear increase with $I$ below 4 mA, consistent with the current dependent $R_{\text{xx}}^{2\omega}$. Subsequently, $R_{\text{xx}}^{2\omega}$ decreases when applying larger current, reflecting the suppression of superconductivity due to the larger current density. These characteristics indicate that the nonreciprocal signal satisfies the relation of $R_{xx}^{2\omega } \propto BI$ in superconducting transition region.

\begin{figure}[!htbp]
	\includegraphics[width=0.9\linewidth,clip]{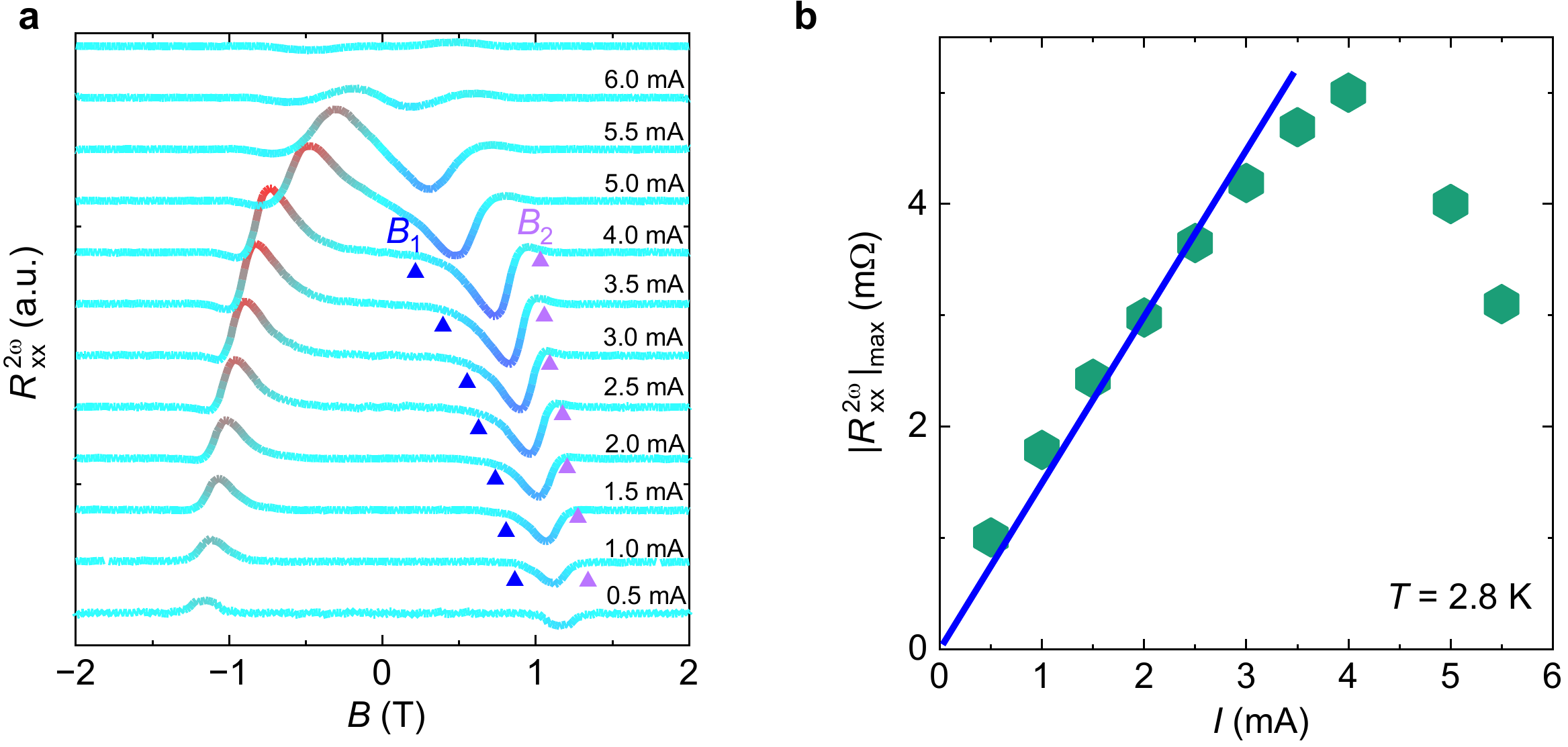}
	\caption{(color online). \textbf{Nonreciprocal charge transport as a function of magnetic field in different current in Ti$_2$O$_3$/GaN heterostructure.} (a) Second harmonic signal as a function of magnetic field in different current range from 0.5  to 6 mA, which represent antisymmetric sign at $I$ \textless 6 mA.  (b) Maximum value of $R_{\text{xx}}^{2\omega}$ at $T$=2.8 K as a function of current $I$, which is extracted from (a). Blue solid line indicates the linear increase as a function of $I$ below 4 mA.
	}
\end{figure}

The red and blue lines in Fig. 4(a) represent second harmonic magnetoresistances with $B$ parallel to the $z$-axis and $y$-axis, respectively, as schematically illustrated inset of the figure. The nonreciprocal response is almost imperceptible when the magnetic field is aligned with the $z$-axis. The results effectively eliminate the possibility of out-of-plane Meissner screening and vortex pinning \cite{Le,Suri,Gutfreund}. The Meissner effect induces two non-dissipative shielding current densities on both sides when the magnetic field is aligned along the $z$-axis. The shielding currents are either added to or subtracted from the applied current on the opposite side, resulting in a corrected measured value. Such extrinsic effect was observed in another sample in Supplementary Materials S11.

Surprisingly, angular dependence of $R_{\text{xx}}^{2\omega}$ reveals two sharp spikes instead of a sinusoidal curve when magnetic field rotating $yz$-plane in Fig. 4(b).  The angle $\varphi$ is defined as the angle in the $yz$-plane, measured from the $z$-axis, as illustrated inset of Fig. 4(b). The spikes located exactly at the position where the $R_{\text{xx}}^{2\omega}$ reaches minimum. We conducted measurements of the magnetic field dependence of the second harmonic magnetoresistance within the range of $\pm 1.5^{\circ}$  near $\varphi=90^{\circ}$, as shown in Fig. 4(c). The peaks of $R_{\text{xx}}^{2\omega}$ were observed when $\varphi$ range from $89.7^{\circ}$ to $90.3^{\circ}$. Notably, the signal disappeared completely when $\varphi=89^{\circ}$ and $\varphi=91^{\circ}$. We also measured the angular dependence of magnetoresistance in the $xz$-plane, with $\theta$  defined as the angle in the $xz$-plane measured from the $z$-axis, as depicted inset of Fig. 4(d). The angular-dependent  $R_{\text{xx}}^{\omega}$ exhibited a two-fold symmetry, with its maximum and minimum occurring at $0^{\circ}$ and $90^{\circ}$. However, the  $R_{\text{xx}}^{2\omega}$ signal disappears completely, suggesting the magnetic field in $xz$-plane does not contribute to the nonreciprocity but only superconductivity suppression.

Together with Fig.S8 and Fig.S9, the angular dependence of  $R_{\text{xx}}^{2\omega}$ seems violated the relation of $R_{\text{xx}}^{2\omega}$=\emph{R}$_0$$\gamma$(\textbf{\emph{B}}$\times$\textbf{\emph{P}})$\cdot$\textbf{\emph{I}}.  For the helical superconductivity, the effective in-plane component of the magnetic fields contribute to the NSC, resulting in a sinusoidal relation for angular-dependent $R_{\text{xx}}^{2\omega}$($\theta$). Our present results, however, an extremely sensitive angular-dependent $R_{\text{xx}}^{2\omega}$($\theta$) suggests a crossover to a symmetric state in the presence of out-of-plane magnetic field. We attribute the crossover to a competition between the depairing current and the vortex depinning current. We observed a violation of the 2D Tinkham-like angular dependence of critical fields from 89 degrees to 91 degrees, where the critical fields are slightly less than those predicted by the 2D Tinkham formula (Fig. S10). When the angle is tilted, out-of-plane vortices enter the device, causing the depairing critical current to be replaced by the vortex depinning current. As the vortex motion does not contribute to nonreciprocity, as shown in Fig. 4(a), the sinusoidal relation is violated.

The nonreciprocity of the depairing critical current is a direct consequence of finite momentum pairing. As the time-reversal symmetry is conserved in our devices, the helical superconductivity is more likely other than FFLO state or chiral superconductivity. However, more experiments, such as Josephson effect with a BCS superconductor, are needed to further clarify the properties of the helical superconductivity.

\begin{figure*}[!htbp]
\includegraphics[width=0.9\textwidth,clip]{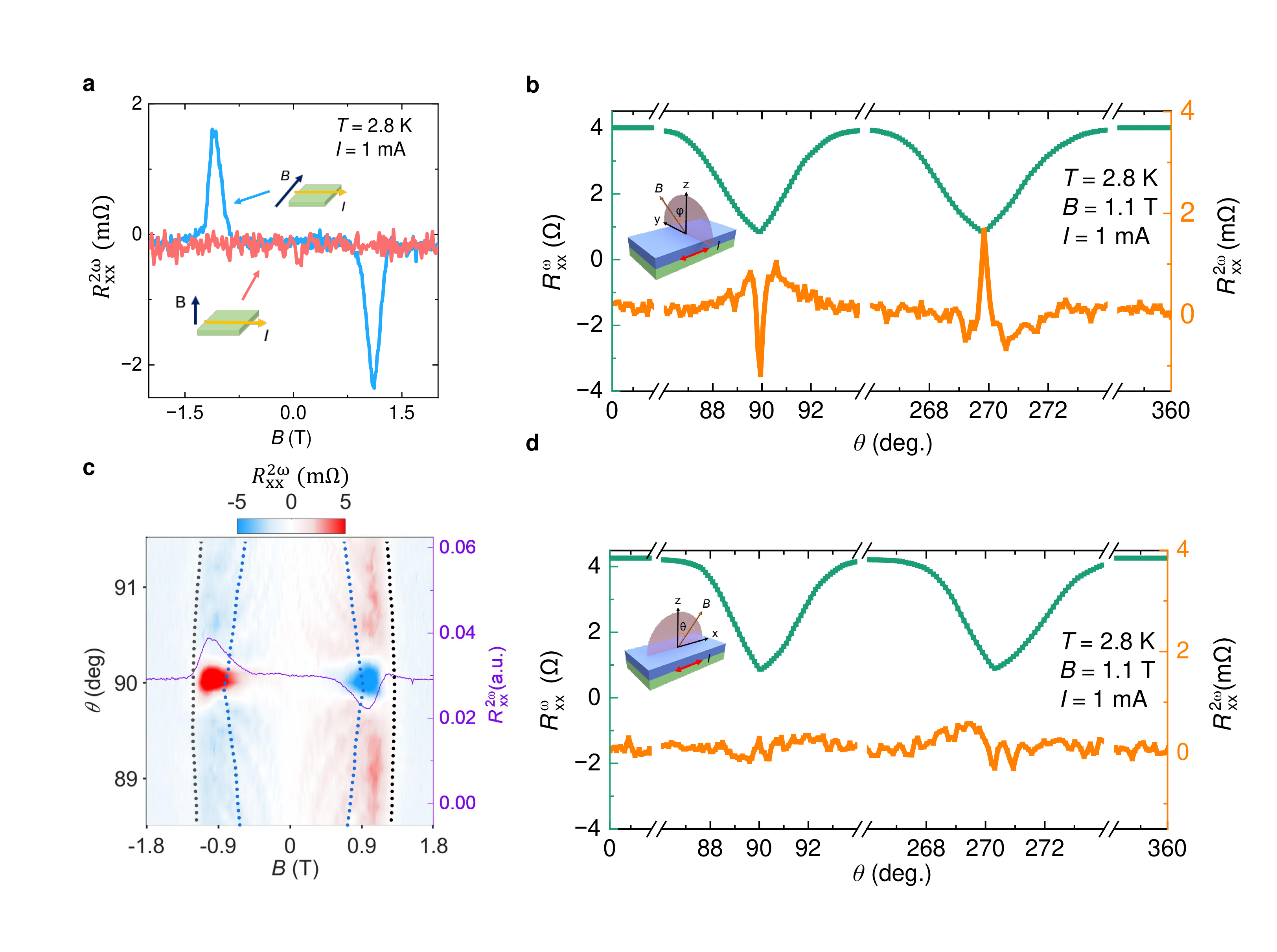}
\caption{(color online). \textbf{Angular-dependent the nonreciprocal charge transport under the alternating current in the condition of 1.1 T and 2.8 K.} (a) The field dependent $R_{\text{xx}}^{2\omega}$ represents strong anisotropy when direction of magnetic field is applied along $y$-axis and $z$- axis. (b) The angular-dependent $R_{\text{xx}}^{\omega}$ and $R_{\text{xx}}^{2\omega}$ by rotating magnetic fields along $yz$-plane. The $R_{\text{xx}}^{\omega}$($\theta$) reveals a two-fold symmetry with maximum and minimum at $0^{\circ}$ and $90^{\circ}$ (green line). And $R_{\text{xx}}^{2\omega}$($\theta$) (orange line) shows a strange anisotropy as well, which exist in an extremely narrow angle region of about $\pm0.5^{\circ}$, and represent antisymmetric sign reversal from 90$^{\circ}$ to 270$^{\circ}$. (c) Mapping of field- and angular-dependent $R_{\text{xx}}^{2\omega}$ in the condition of 2.8 K, where field is rotated within $yz$-plane. $R_{\text{xx}}^{2\omega}$ exists within a narrow angle region of $\pm$0.5$^{\circ}$, which is consistent with Fig 4(b). (d) Angular-dependent $R_{\text{xx}}^{\omega}$ and $R_{\text{xx}}^{2\omega}$ by rotating field within $xz$-plane. $R_{\text{xx}}^{\omega}$ indicates a normal superconductivity suppression from effective field, while $R_{\text{xx}}^{2\omega}$ is absent.
}
\end{figure*}

In summary, our investigation focused on understanding the nonreciprocity observed in the interfacial superconductor. The temperature and current dependence of $R_{\text{xx}}^{2\omega}$ indicate that the nonreciprocal strength $\gamma$ (12 $\text{A}^{-1}\text{T}^{-1}$) is greater than that of normal metals and semiconductors. Furthermore, the temperature and current dependence of $R_{\text{xx}}^{2\omega}$ exhibit similar trends, wherein an initial increase is followed by a subsequent decrease as the temperature or current rises. The angular dependence of $R_{\text{xx}}^{2\omega}$ reveals that the spikes, which coincide with the point where $R_{\text{xx}}^{\omega}$ reaches its minimum, are associated with two-dimensional superconductivity. It is likely that the symmetry breaking is a consequence of unconventional pairing.

This research was supported in part by the Ministry of Science and Technology (MOST) of China (No. 2022YFA1603903), the National Natural Science Foundation of China (Grants No. 12004251, 12104302, 12104303, 12304217), the Science and Technology Commission of Shanghai Municipality, the Shanghai Sailing Program (Grant No. 21YF1429200), the Interdisciplinary Program of Wuhan National High Magnetic Field Center (WHMFC202124), the open project from State Key Laboratory of Surface Physics and Department of Physics, Fudan University (Grant No. KF2022-13), the Shanghai Sailing Program (23YF1426900).

%

\end{document}